\documentclass[a4paper,fleqn,usenatbib]{mnras}

\usepackage[T1]{fontenc}
\usepackage{ae,aecompl}

\usepackage{graphicx}	
\usepackage{amsmath}	
\usepackage{amssymb}	

\usepackage{txfonts}
\usepackage{hyperref}
\usepackage{longtable}

\def\ms{\hbox{\,m\,s$^{-1}$}}         
\def\m2s2{\hbox{\,m$^{2}$\,s$^{-2}$}} 
\def\kms{\hbox{\,km\,s$^{-1}$}}       
\def\Msun{\hbox{$\mathrm{M}_{\odot}$}}             
\def\Rsun{\hbox{$\mathrm{R}_{\odot}$}}

\def \kepler{\emph{Kepler}}

\def\kepler{\emph{Kepler}}

\def\logg{$\log g$}
\def \kms{km\,s$^{-1}$}
\def \1s{$1\,\sigma$}

\def \t0{T$_0$}

\def\Msun{\hbox{$\mathrm{M}_{\odot}$}}             
\def\Rsun{\hbox{$\mathrm{R}_{\odot}$}}

\def\e350{K2-19}

\def\ten[#1]{$\;\times 10^{#1}$}

\newcommand{\degree}{\ensuremath{^\circ}}

\newcommand{\mearth}{{\hbox{$M_{\oplus}$}}}
\newcommand{\rearth}{{\hbox{$R_{\oplus}$}}}

\title[TTVs of  K2-19]{Photo-dynamical mass determination of the multi-planetary system K2-19\thanks{Based on observations collected  with the NASA \kepler\ satellite and with the {\it SOPHIE}  spectrograph on the 1.93-m telescope at Observatoire de Haute-Provence (CNRS), France.}}           
\author[S.~C.~C.~Barros et al.]{S.~C.~C.~Barros$^{1,2}$\thanks{E-mail: susana.barros@astro.up.pt}, J.~M. Almenara$^{3,4}$,  O.~ Demangeon$^{1}$, M.~Tsantaki$^{2}$, A.~Santerne$^{2}$, \and
D.~J.~Armstrong$^{5}$,   D.~Barrado$^{6}$, D.~Brown$^{5}$,  M.~Deleuil$^{1}$,  J.~Lillo-Box$^{6}$, H.~Osborn$^{5}$,   \and D.~Pollacco$^{5}$, L.~Abe$^{7}$, P.~Andre$^{8}$, P.~Bendjoya$^{7}$,  I.~Boisse$^{1}$, A.~S.~Bonomo$^{9}$, F.~Bouchy$^{1}$,   \and G.~Bruno$^{1}$,  J.~Rey Cerda$^{10}$, B.~Courcol$^{1}$, R. F.~D\'iaz$^{10}$,  G.~ H\'ebrard$^{11,12}$, J.~Kirk$^{5}$,   \and J.C.~Lachuri\'e$^{8}$,   K.~W.~F.~Lam$^{5}$, P.~Martinez$^{8}$, J.~McCormac$^{5}$,  C.~Moutou$^{13,1}$, A.~Rajpurohit$^{1}$,   \and J.-P.~Rivet$^{7}$,  J.~Spake$^{5}$, O.~Suarez$^{7}$, D. Toublanc$^{8,14}$ and S.~R.~Walker$^{5}$ \\
$^{1}$Aix Marseille Universit\'e, CNRS, LAM (Laboratoire d'Astrophysique de Marseille) UMR 7326, 13388, Marseille, France\\
$^{2}$Instituto de Astrof\'isica e Ci\^encias do Espa\c{c}o, Universidade do Porto, CAUP, Rua das Estrelas, PT4150-762 Porto, Portugal\\
$^{3}$Univ. Grenoble Alpes, IPAG, F-38000 Grenoble, France\\
$^{4}$CNRS, IPAG, F-38000 Grenoble, France\\
$^{5}$Department of Physics, University of Warwick, Gibbet Hill Road, Coventry, CV4 7AL, UK\\
$^{6}$Departamento de Astrof\'isica, Centro de Astrobiolog\'ia (CSIC-INTA), ESAC campus 28691 Villanueva de la Ca\~nada (Madrid),Spain \\
$^{7}$Laboratoire Lagrange, UMR7239, Universit\'e de Nice Sophia-Antipolis, CNRS, Observatoire de la Cote d'Azur, F-06300 Nice, France \\
$^{8}$Observatoire de Belesta en Lauragais - Assoc. Astronomie Adagio 30 Route de Revel 31450 Varennes, France\\
$^{9}$INAF - Osservatorio Astrofisico di Torino, via Osservatorio 20, 10025, Pino Torinese, Italy\\
$^{10}$Observatoire Astronomique de l'Universite de Gen\`eve, 51 chemin des Maillettes, 1290 Versoix, Switzerland\\
$^{11}$Institut d'Astrophysique de Paris, UMR7095 CNRS, Universit\'e Pierre \& Marie Curie, 98bis boulevard Arago, 75014 Paris, France\\
$^{12}$Observatoire de Haute-Provence, Universit\'e d'Aix-Marseille \& CNRS, 04870 Saint Michel l'Observatoire, France\\
$^{13}$CNRS, Canada-France-Hawaii Telescope Corporation, 65-1238 Mamalahoa Hwy., Kamuela, HI 96743, USA\\
$^{14}$Universite de Toulouse, UPS-CNRS, IRAP, 9 Av. colonel Roche, 31028 Toulouse cedex 4, France }

\date{Accepted XXX. Received YYY; in original form ZZZ}

\pubyear{2015}

\begin{document}
\label{firstpage}
\pagerange{\pageref{firstpage}--\pageref{lastpage}}
\maketitle

\begin{abstract}
\e350 is the second multi-planetary system discovered with K2 observations. The system is composed of two Neptune size planets close to the 3:2 mean-motion resonance. 
To better characterise the system we obtained two additional transit observations of \e350b and five additional  radial velocity observations. These were combined with K2 data and fitted simultaneously with the system dynamics (photo-dynamical model) which increases the precision of the transit time measurements. The higher transit time precision allows us to detect the chopping signal of the dynamic interaction of the planets that in turn permits to uniquely characterise the system.  
Although the reflex motion of the star was not detected, dynamic modelling of the system allowed us to derive planetary masses of $M_b= 44 \pm 12\,$ \mearth\ and  $M_c = 15.9 \pm 7.0\,$ \mearth\ for the inner and the outer planets respectively, leading to densities close to Uranus. We also show that our method allows the derivation of mass ratios using only the 80 days of observations during the first campaign of K2.
\end{abstract}




\section{Introduction}

 Transit timing variations (TTVs) are caused by the mutual gravitational interaction of planets which perturb each others' orbit. These are larger when the planets are close to mean-motion resonances  \cite[MMRs;][]{Escude2002,Holman2005,Agol2005}. Therefore, most of the known TTV systems are close to MMRs. For example, the first detected system showing TTVs,  Kepler-9 \citep{Holman2010}, is a system composed of a pair of transiting Saturn-mass planets near the 2:1 MMR and an inner earth-sized companion. However, the high precision and long baseline of  \kepler\ \citep{Borucki2010} also made it possible to detect TTVs in systems away from resonance \citep{Bruno2015, Almenara2015}.

In near-resonant systems the resonant angles which measure the displacement of the longitude of the conjunction from the periapsis of each planet, circulate (or librate) over a period much longer than the orbital period of the outer planet, called the libration period or super period. \citet{Lithwick2012} showed that for systems close to first order MMRs the TTV signal is sinusoidal and that this libration period is inversely proportional to the distance to the resonance. Furthermore, the TTV amplitude depends on planet mass and hence TTVs provide another method to estimate planetary masses.
  However, since the TTV amplitude also depends on the free eccentricity, the TTV inversion problem is degenerate \citep{Lithwick2012} unless the TTV curves are known with high accuracy as in the case of Kepler-9 \citep{Holman2010,Dreizler2014,Deck2015}.  

Besides the resonant long term periodicity, TTVs show a short-timescale component ``chopping'' which is related to the closest approach of both planets when the mutual gravitation interaction is strongest and occurs in the synodic timescale \citep{Nesvorny2010, Nesvorny2014, Deck2015}. The chopping component of the TTVs is smaller than the resonant term and hence more difficult to detect. Recently, \citet{Nesvorny2014, Deck2015} showed that for low eccentricity, nearly co-planar systems the synodic chopping term depends only on the mass ratios and not on the eccentricity and hence makes it possible to uniquely estimate the planetary masses. The detection of the chopping component of the TTVs allowed the determination of unique solutions for a few systems e.g. KOI-872 and KOI-142 \citep{Nesvorny2012,Nesvorny2013}.

Dynamic analyses of TTVs in \kepler\ transiting multi-planetary systems have allowed a better characterisation of the system and/or helped confirm the planetary nature of many candidates. 
In particular, \kepler\ data has shown a pile-up of planet pairs wide but close to first order MMRs and a gap narrow of these resonances \citep{Lissauer2011a, Fabrycky2012}. This could be due to dissipation leading to ``resonant repulsion'' \citep{Lithwick2012b, Batygin2013}.

After the failure of two out of four of the reaction wheels of the \kepler\ satellite the pointing accuracy was severely degraded. Cleaver engineering allowed the continuation of the mission in a new configuration named K2 \citep{Howell2014}. K2 observes 4 fields a year close to the Ecliptic. The short duration of the observations of each field ($\sim 80\,$ days) does not favour the detection of TTVs amongst planetary candidates discovered in these observations.

 \e350 (EPIC201505350) is a multi-planetary system detected in the K2 Campaign 1 (C1) data by \citet{Armstrong2015}. The K2 observations show 2 transiting planets one with an orbital period  $P_b \sim 7.92\,$days and radius $R_b$=$7.23 \pm 0.60\, $\rearth\ and a companion close to the 3:2 MMR with an orbital period  $P_c \sim 11.91\,$days and radius $R_b$ = $4.21 \pm 0.31\, $\rearth. The closeness to resonance implied that  \e350 was a good candidate for TTV and the brightness of the host star allowed follow-up transit observations from the ground. Approximately 200~days after the end of the K2 C1, a ground based transit was obtained showing TTVs of the inner planet with an amplitude of 1 hour, allowing the authors to validate the system.

In this paper we present a photo-dynamical analysis of the \e350 system using K2 observations, three ground based transits of \e350b and radial velocities of the host star observed with the SOPHIE spectrograph.
We show that although the libration period is not well constrained,  we are able to characterise the system due to the detection of the short period TTVs (chopping) in both companions. This was possible due to a photo-dynamical model that simultaneously fits the photometric transit observations, the radial velocities and the system dynamics. In Section~2, we describe the observations and in Section~3, we present the spectral characterisation of the host star. We describe the photo-dynamic model in Section~4 and present our results in Section~5. Finally, we discuss our results in Section~6.

\section{Observations}

\subsection{K2 data}

\e350 was observed during Campaign 1 of the K2 mission between  2014 June 3 and 2014 August 20 spanning $\sim 80\,$ days. We downloaded the pixel data from the Mikulski Archive for Space Telescopes (MAST)\footnotemark \footnotetext{$http://archive.stsci.edu/kepler/data\_search/search.php$}
 and used a modified version of the CoRoT imagette pipeline \citep{Barros2014} to extract the light curve. We computed an optimal aperture based on signal-to-noise of each pixel. The background was estimated using the $3\sigma$ clipped median of all the pixels in the image and removed before aperture photometry was performed. We also calculated the centroid using the Modified Moment Method by \citet{Stone1989}. For \e350, we found that a 30 pixels photometric aperture resulted in the best photometric precision, this is roughly equivalent to an aperture radius of 3 pixels. A full description of the pipeline will be given in Barros et al (in prep).

The degraded pointing stability of the K2 mission couples with pixel sensitivity variations to introduce systematics in the raw light curves. Several methods to correct the systematics have been applied to the K2 data \citep{Vanderburg2014, Aigrain2015, Foreman-Mackey2015,Armstrong2015b, Lund2015}. In our case to correct for this flux dependence with position we used a procedure similar to \citet{Vanderburg2014}. Due to the poor pointing accuracy, the targets slowly drift in the CCD and to correct for this, every 6 hours the spacecraft thursters are fired returning the target to the initial position.  \citet{Vanderburg2014} showed that the movement of the satellite  was mainly in one direction. Thus, for each roll of the spacecraft, the target crosses a similar path in the CCD. This allows to the use self-flat-fielding to correct the flux position variation by calculating the mean flux at each of a series of centroid position bins.

Following,  \citet{Vanderburg2014}, we start by estimating and removing stellar activity with a spline filter. Then we calculate the main direction of motion using principal component analysis. Finally, the flux dependence with position is computed and the correction is applied to the data.  This 1D approximation starts failing after  $\sim 10\,$days due to an extra slow drift of the satellite along the direction perpendicular to the main rolling motion. Hence, to maintain the 1D approximation, the light curves are divided in 8 segments of equal duration and for each segment we performed the decorrelation method described above. After this self-flat fielding procedure the spline filter is re-added to the light curve to avoid affecting the transits shape.

The final light curve of \e350, with a mean RMS of $840\,$ ppm  contains ten transits of \e350b and seven of \e350c. Three of the transits are simultaneous for the two planets. Noteworthy, the star shows rotational variability with a peak-to-peak amplitude of 1.5\% during the K2 observations. For the transit analysis we extracted individual transit light curves with length corresponding to three transit duration's and centred at the mid-transit time. Each transit was normalised by a quadratic baseline function. 
 The K2 transit observations of both planets are shown in the top and middle panel of Figure~\ref{fig.transit}. We also show the median value of the distribution of models described in Section~\ref{modeling}.

\subsection{Ground based transits}

The brightness of the host star (Kepmag = 12.8 mag) allows further observations from the ground. Hence, after the candidate detection we initiated a follow-up campaign.
Three transits of \e350b were observed from the ground $198\,$days after the K2 observations.

\begin{figure*}
\centering
\includegraphics[width=1.9\columnwidth]{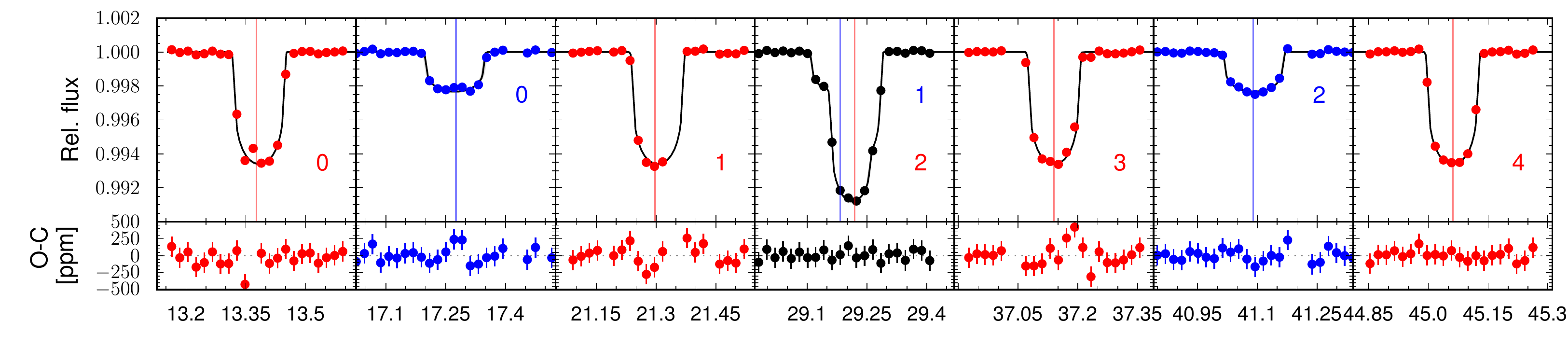}\\
\includegraphics[width=1.9\columnwidth]{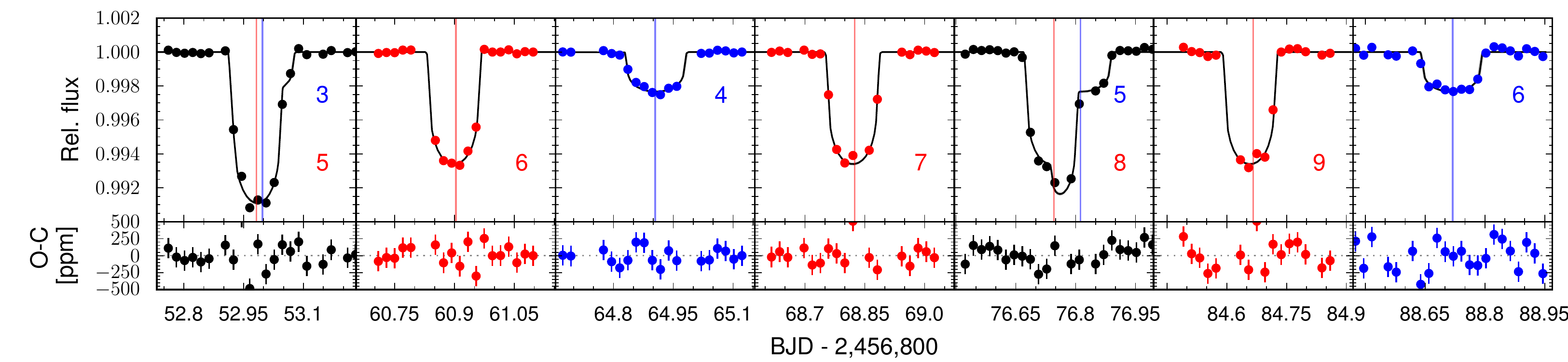}\\
\includegraphics[width=1.9\columnwidth]{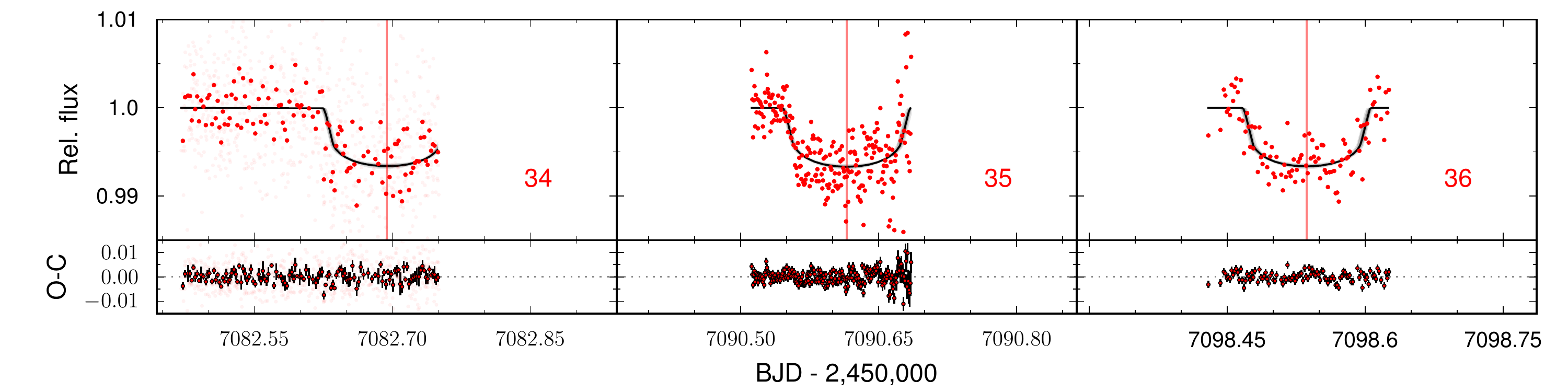}
\caption{Transit observations of \e350 in order of the transit time. For each transit we overploted the transit epoch and the transit time expected by a linear ephemeris (vertical lines). The black curve is the median value of the distribution of models (presented in Section~\ref{modeling}) corresponding to 1,000 random MCMC steps.}
\label{fig.transit}
\end{figure*}

The first was observed on the 2015 February 28 by the 0.4-m Near Infrared Transiting ExoplanetS (NITES) telescope \citep{McCormac2014} (Figure~\ref{fig.transit}, bottom panel, epoch = 34). The telescope was defocused and a $20\,$second exposure time was used. Observations were performed with no filter. The data reduction was performed using standard IRAF\footnotemark \footnotetext{IRAF is distributed by the National Optical Astronomy Observatories, which are operated by the Association of Universities for Research in Astronomy, Inc., under cooperative agreement with the National Science Foundation.} routines and DAOPHOT \citep{Stetson1987} to perform aperture photometry. This transit was already presented in \citet{Armstrong2015}.

The second transit was observed on 2015 March 8 at the $1-$m C2PU/Omicron telescope in Calern (Observartoire de la Cote d'Azur). It is shown in the bottom panel of Figure~\ref{fig.transit} (Epoch $= 35$).
The exposure time was $60\,$seconds and the Johnson-R filter was used. The data was reduced using the astro-ImageJ aperture photometry tool.

The third transit was observed on the 2015 March 16 at the Belesta $82-$cm telescope (Figure~\ref{fig.transit}, bottom panel, epoch = 36).
The exposure time was $120\,$seconds and Johnson-R was used. The data was reduced with the ATV IDL tool \citep{Barth2001} which performs aperture photometry.

For each observation, differential photometry was performed using a careful selection of reference stars. The times were converted to Barycentric Dynamical Time (TDB) using the 
IDL codes kindly made available by \citet{Eastman2010}.

\subsection{Spectroscopic observations}

We obtained 10 spectroscopic observations of \e350 from 2015 February 21 to 2015 April 25 with the SOPHIE spectrograph mounted on the 1.93m telescope at the Observatoire de Haute-Provence \citep{Perruchot2011, Bouchy2013}. SOPHIE is a thermally stable high-resolution echelle optical spectrograph fed by a fibre link from the Cassegrain focus of the telescope. The fibre has a diameter of 3" on sky. The observations were obtained in the high-efficiency (HE) mode, which has a resolution R$\sim40\, 000$ and covers a wavelength range of 390-687 nm.

The radial velocities were derived with the SOPHIE pipeline by computing the weighted cross-correlation function (CCF) \citep{Baranne1996, Pepe2002} with a G2 mask. The extracted RVs include corrections for charge transfer inefficiency of the SOPHIE CCD \citep{Bouchy2009} using the procedure described in \citet{Santerne2012}. The SOPHIE HE mode exhibits instrumental variations at long time-scales with an amplitude of a few \ms. These were corrected for using observations of a bright and stable star, HD\,185144, obtained on the same nights and with the same instrument setup \citep{Santerne2014}. The bisector (Vspan) and FWHM errors have been estimated using the scaling relation published in \citet{Santerne2015}.

\begin{figure}
\centering
\includegraphics[width=0.45\textwidth]{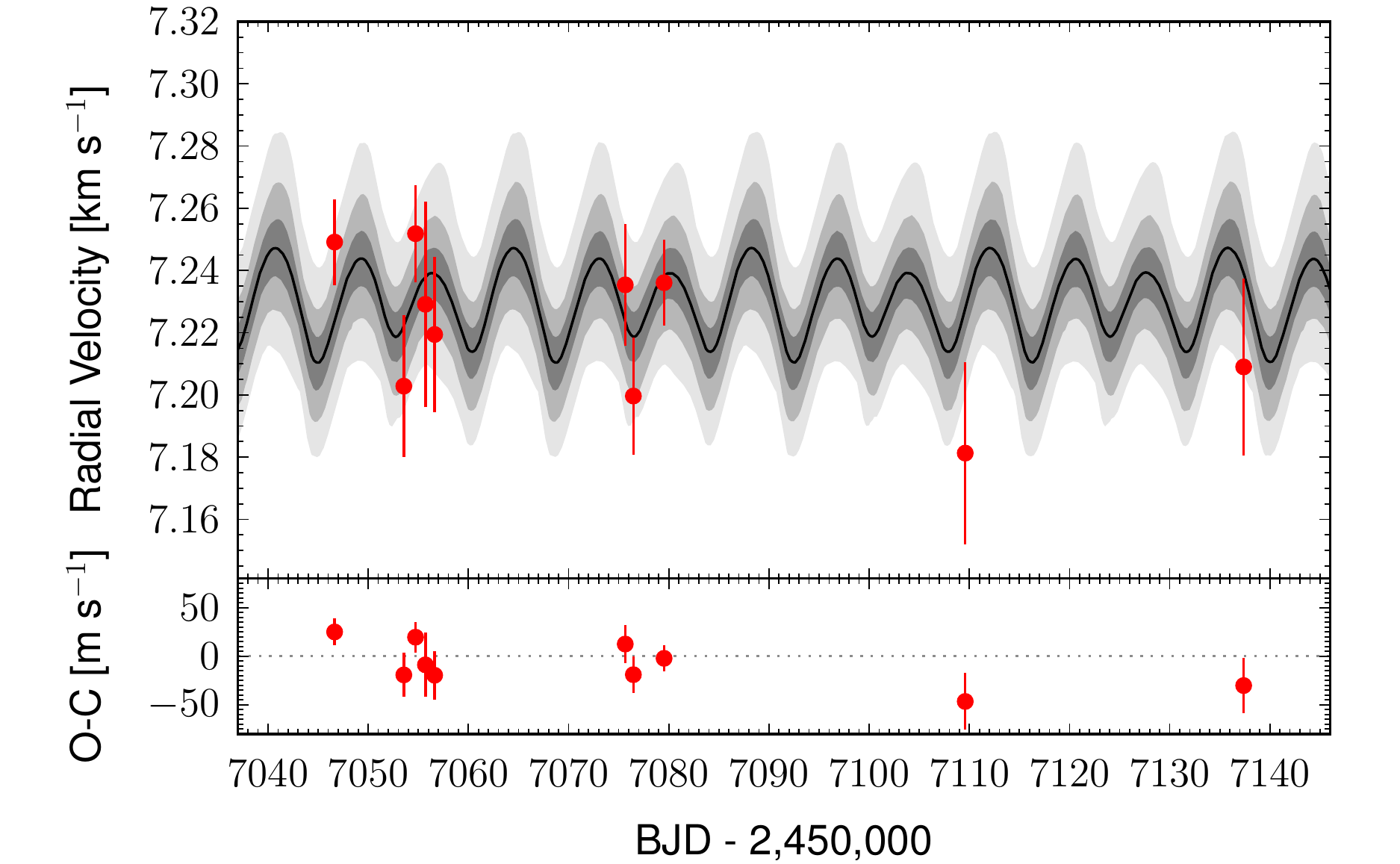}
\caption{SOPHIE radial velocities as a function of time and the corresponding residuals. The over-plotted black curve is the median value of the distribution of models  (described in Section~\ref{modeling}) corresponding to 1,000 random MCMC steps, and the different shades of grey represent the 68.3, 95.5, and 99.7\% Bayesian confidence intervals. \label{fig.RVdata}}
\end{figure}

The average signal-to-noise ratio of each spectra is 18 per pixel at 5500 \AA\ and the average RV uncertainty is 22~\ms. 
The measurements and the respective uncertainties are given in Table~\ref{table:rv}. In the same table we also list the exposure time, signal-to-noise ratio per pixel at 550 nm and the bisector span of the cross-correlation function. The radial velocity measurements are shown in Figure~\ref{fig.RVdata} together with the model described in Section~\ref{modeling}.

\begin{table*}
\centering
\caption{Radial velocity measurements for \e350 taken with SOPHIE.  \label{table:rv}}
\setlength{\tabcolsep}{4mm}
\begin{minipage}[c]{2\columnwidth} 
\centering
\begin{tabular}{l l l c c c c c c}
\hline
\hline
$\rm BJD_{UTC}$     & RV          & $\sigma_{RV}$ & Vspan  & $\sigma_{Vspan}$ & FWHM &  $\sigma_{FWHM}$ &T$_{exp}$ & S/N/pix    \\
-2457000  &(\kms) & (\kms) & (\kms) & (\kms) & & & (s)       & (550 nm)\\
\hline
046.64374   &    7.250   &   0.014   &   -0.057   &   0.036   &    10.406   &   0.034   &    2700   &   21.6 \\
053.55195   &    7.203   &   0.023   &   -0.022   &   0.060   &    10.537   &   0.057   &    1800   &   19.2 \\
054.72060   &    7.252   &   0.016   &   0.0002   &   0.041   &    10.601   &   0.039  &    2200   &   22.1 \\
055.70836   &    7.229   &   0.033   &   -0.026   &   0.087   &    10.587   &   0.083   &    400   &   13.5 \\
056.62866   &    7.219   &   0.025   &   -0.091   &   0.065   &    10.303   &   0.062   &    1573   &   16.7 \\
075.65385   &    7.235   &   0.020   &   -0.013   &   0.055   &    10.265   &   0.049   &    3600   &   17.2 \\
076.46383   &    7.200   &   0.019   &   -0.016   &   0.049   &    10.525   &   0.047   &    3600   &   18.7 \\
079.53034   &    7.236   &   0.014   &   0.001   &   0.036   &    10.335   &   0.034  &    3600   &   26.8 \\
109.59053   &    7.181   &   0.030   &   -0.102   &   0.076   &    10.366   &   0.073   &    3600   &   14.1 \\
137.38239   &    7.209   &   0.028   &   -0.103   &   0.074   &    10.573   &   0.071   &    3066   &   14.5 \\
\hline
\end{tabular}
\end{minipage}
\end{table*}

\section{Host star \label{sec.stellarparams}}

We obtained the atmospheric parameters of the host star from the spectral analysis of ten co-added
SOPHIE spectra. First, we subtracted any sky contamination (using the spectra of fibre B) 
from the spectra pointing to the source (in fibre A) while correcting for the relative efficiency 
of the two fibres. The final spectrum has a S/N of $\sim$35 around 6070\AA{}.

To derive the atmospheric parameters, namely the effective temperature ($T_{\mathrm{eff}}$), surface gravity ($\log g$), 
metallicity ($[Fe/H]$), and microturbulence ($\xi_{t}$), we followed the methodology described in \cite{Tsantaki2013}.
This method relies on the measurement of the equivalent widths (EW) of Fe\,{\sc i}/Fe\,{\sc ii} lines and by imposing 
excitation and ionisation equilibrium. The analysis was performed in local thermodynamic equilibrium using a grid of 
model atmospheres \citep[ATLAS9,][]{Kurucz1993} and the radiative transfer code MOOG \citep{Sneden1973}. 
Due to the low S/N of our spectrum, the EWs were derived manually using the IRAF \texttt{splot} task to avoid errors in the measurements. 

From the above analysis, we obtain the following atmospheric parameters for the host star: $T_{\mathrm{eff}}$ = 5390 $\pm$ 180\,K, 
$\log g$ = 4.42 $\pm$ 0.34\,dex, $\xi_{t}$ = 1.02 $\pm$ 0.24\,km\,s$^{-1}$, and  $[Fe/H]$ = 0.19 $\pm$ 0.12\,dex, hence it is a K-dwarf. 
The spectroscopic gravity of the EW method is usually not well constrained when compared with other more 
model-independent methods such as from asteroseismology \citep{Mortier2014} or from the parallax \citep{Tsantaki2013, Bensby2014}. Therefore, we calculated the correction for the spectroscopic 
surface gravity using the calibration described in \cite{Mortier2014} (Eq.\,2) and found perfect agreement with our spectroscopic results 
($\log g_{corr}$ = 4.42\,dex).  The derived parameters agree with the ones previously presented by \citep{Armstrong2015}. 
The larger number of spectra led to a lower uncertainty of the combined spectra which allowed a much better constraint on the derived stellar spectroscopic parameters. The uncertainty on these are approximately half of those found previously.

Using the above parameters for the host star $T_{\mathrm{eff}}$, $\log g$ and  $[Fe/H]$  we derive the stellar mass and radius by interpolating the stellar evolution models of Geneva \citep{Mowlavi2012}, Dartmouth \citep{Dotter2008} and PARSEC \citep{Bressan2012} using the MCMC described in \citet{Diaz2014}. We obtained $ M_*$ = $0.918_{-0.070}^{+0.086}$ \Msun\ and $  R_*$ = $0.926_{-0.069}^{+0.19} $  \Rsun. These are also in agreement with the values presented by \citet{Armstrong2015}.

\section{Photo-dynamical model}

\label{modeling}
All the transits and radial velocities were modelled simultaneously with an n-body dynamical integrator that accounts for the gravitational interactions between all components of the system. We use the \textsc{mercury} n-body integrator \citep{Chambers1999} to compute the 3 dimensional position and velocity of all system components as a function of time. We assume that only the host star and two planets are present. The stellar velocity projected onto the line-of-sight is used to  model the observed radial velocities. To model the transits, we use the \citet{Mandel2002} transit model parametrised by the planet-to-star radius ratio,  quadratic limb darkening coefficients for each filter and using the sky projection of the planet-star separation computed from the output of \textsc{mercury}.  To account for the 29.4 minutes integration time, the transit model was over-sampled by a factor of 20 and binned to the cadence of the data points.  This photo-dynamical model is coupled to a Monte Carlo Markov Chain (MCMC) routine, described in detail in \citet{Diaz2014}, in order to derive the posterior distribution of the parameters.

For each step of the MCMC a \textsc{mercury} run is performed using Bulirsch-Stoer algorithm with a $0.01\,$day time step for the photometry (that implies a maximum model related photometric error of 1ppm) and a $0.02\,$day time step for the radial velocities. The parameters of the system used in the fit are the stellar mass and radius, the limb darkening coefficients, the systemic velocity, the planetary masses, the planet-to-star radius ratio, and the planetary orbital parameters (orbital semi-major axis $a$, eccentricity $e$, orbital inclination $i$, argument of the periastron $\omega$, longitude of the ascending node, $n$, and the mean anomaly $M$)  at reference time $t_{\mathrm{ref}}$ which we choose to be $2,456,813$~BJD. To minimise correlations between the model parameters which prevent adequate exploration of the parameter space, we used the \citet{Huber2013} parametrisation (see Table~\ref{mcmc}). Furthermore, we fitted a normalisation factor for each photometric data set and a multiplicative jitter parameter for the radial velocities and each photometric data set.

To define the reference plane we set the longitude of the ascending node of planet b to 180\degree.  The system is symmetric relative to the inclination of one of the planets, hence we constrain the inclination of the planet b to one hemisphere ($i_b$ <90\degree) but allow the planet c to transit both hemispheres  ($i_b$ <180\degree).  As shown by \citet{Almenara2015}, if the radial velocity measurements and the transit photometry have high enough precision, our model completely constrains the system without requiring additional information about the host star. However, we performed test runs and concluded that the radial velocity measurements are not precise enough to detect the stellar reflex velocity due to any of the planets. Moreover, the transit derived stellar density is not well constrained due to the poor sampling of the K2 observations and the precision of the ground based transits. Therefore, to help constrain the system we included Gaussian priors on the stellar mass and radius derived from the spectroscopic observations.We also use Gaussian priors for the limb darkening coefficients based on the tables of \citet{Claret2011}. We use non-informative uniform priors for all the remaining model parameters. We run 46 independent MCMC chains and combined the results as described by \citet{Diaz2014} resulting in a final merged chain with3500 independent points. Further details about the photo-dynamic method can be found in \citet{Almenara2015}.

\section{Results}

\begin{figure*}
\centering
\includegraphics[width=1.1\textwidth]{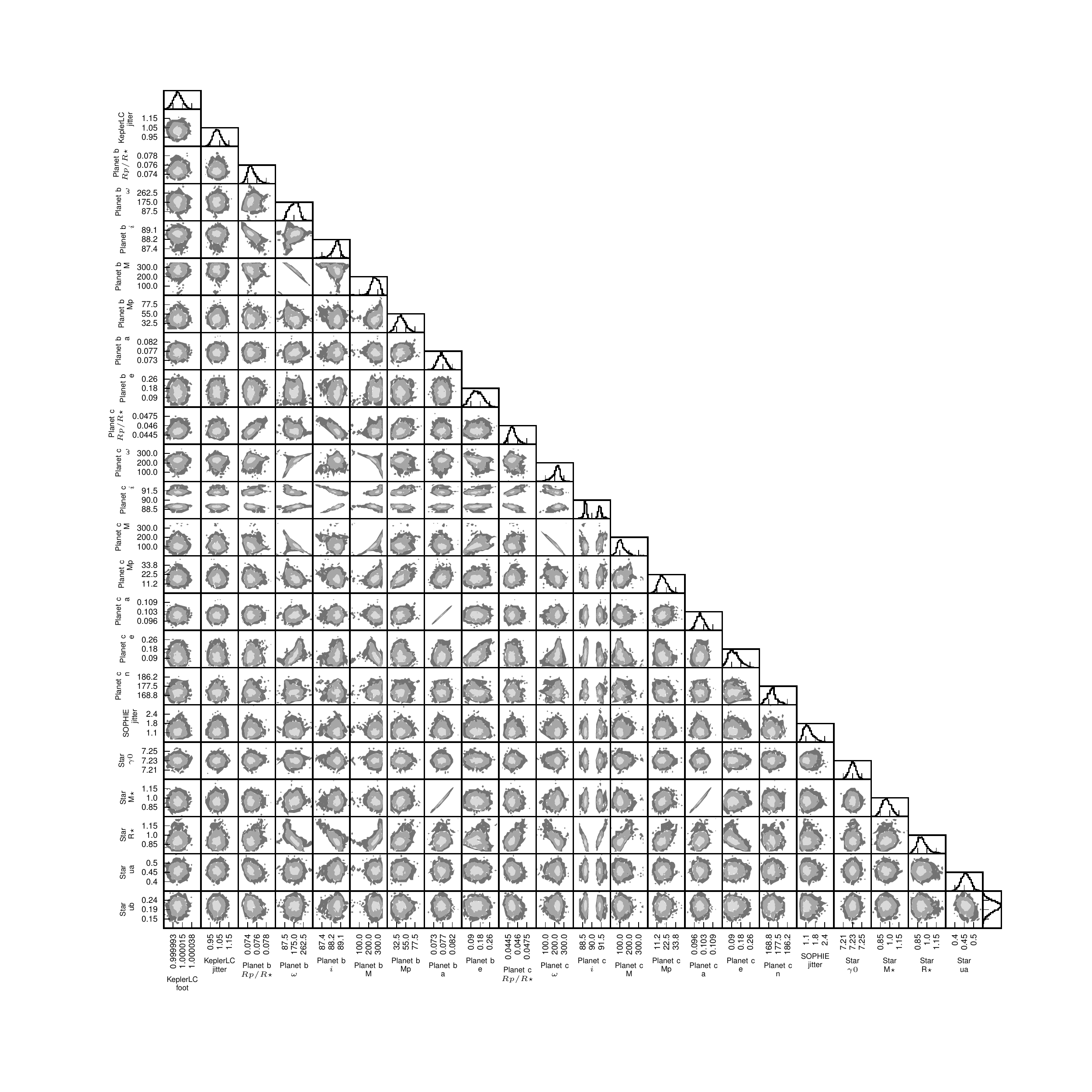}
\caption{Correlation plots and posterior distribution of the fitted parameters of our model. We cannot distinguish which hemisphere planet c orbits as the two configurations are equally probable.   \label{correlation} }
\end{figure*}

 The mode and the 68.3\% confidence interval for the derived system parameters are given in Table~\ref{mcmc}. We find that \e350b has a mass of  $44 \pm 12\,$ \mearth\ and radius of $7.46 \pm 0.76\, $\rearth\ and  \e350c has a mass of 15.9$^{+7.7}_{-2.8}\,$ \mearth\  and a radius of $4.51 \pm 0.47\, $\rearth\ . The inclination of planet c could not be constrained to either hemisphere. For clarity, in Table~\ref{mcmc} we give only the solution where planet c orbits the same hemisphere as planet b. However, both values of the inclination: 88.92$^{+0.14}_{-0.41}$ and $91.19^{+0.41}_{0.14}$ are equally probable. The correlations between the parameters and the posterior parameter distributions are presented in Figure~\ref{correlation}. 

As mentioned above, in our analysis we included stellar priors to scale the system. However, with only photometry and radial velocities it is possible to derive a solution independent of stellar models albeit with a poorer constrain on the scale of the system. We re-run the photo-dynamical model with uniform priors in stellar mass and density and uniform priors in the linear limb darkening coefficients. As expected, this results in larger errors for the physical parameters that depend on the scale such as the stellar parameters: $R_\star$ = $0.73^{+0.98}_{-0.28} \hbox{$\mathrm{R}_{\odot}$}$ and $\rho_{\star}$ = $2.01^{+0.72}_{-0.24} \rho_\odot$ which propagate to other physical parameters. Noteworthy is that the poorer constraint on the stellar density biases the orbital eccentricities to higher values $e_{b}$ = $0.309^{+0.055}_{-0.10}\,$ and $e_{c}$ = $0.271 \pm 0.072\,$  than the values derived when stellar priors are included (see Table~\ref{mcmc}). However, the mass ratios are determined by the system dynamics and hence are independent of the scale of the system. Therefore, they are well constrained and in agreement with the results obtained with the stellar priors $q_+$ = $\frac{M_{p,b}+M_{p,c}}{M_\star}\,\,$ = $0.000142^{+0.000038}_{-0.000023}$ and $q_p = \frac{M_{p,c}}{M_{p,b}}\,\,$ = $0.474^{+0.16}_{-0.084} $ . In case more precise stellar parameters are determined in the future, these mass ratios should be used to re-derive the planetary masses and the radius ratio to re-derive the planetary radius.

\begin{table*}
\renewcommand{\arraystretch}{0.7}
\centering
\caption{Model parameters. Posterior mode and 68.3\% credible intervals. The orbital elements have the origin at the star (Asteroidal parameters in the \textsc{mercury} code) and are given for the reference time $t_{\mathrm{ref}} = 2,456,813$~BJD.}\label{mcmc}
\begin{tabular}{lcc}
\hline
Parameter & \multicolumn{2}{c}{Mode and 68.3\% credible interval} \\
\hline
Stellar mass, $M_\star$ [\Msun]$^{\bullet}$               & 0.949 $\pm$ 0.077 & \\
Stellar radius, $R_\star$ [\Rsun]$^{\bullet}$             & 0.913 $\pm$ 0.094 & \\
Stellar density, $\rho_{\star}$ [$\rho_\odot$]            & 1.25 $\pm$ 0.33   & \\
Surface gravity, \logg\ [cgs]                           & 4.493 $\pm$ 0.085 & \\
Systemic velocity, $\gamma$ [\kms]$^{\bullet}$           & 7.2296 $\pm$ 0.0080 & \\
Linear limb darkening coefficient, $u_a$$^{\bullet,\ast}$ & \multicolumn{2}{l}{0.460 $\pm$ 0.026, 0.442 $\pm$ 0.036, 0.476$^{+0.027}_{-0.045}$, 0.435$^{+0.049}_{-0.023}$  } \\
Quadratic limb darkening coefficient, $u_b$$^{\bullet,\ast}$ & \multicolumn{2}{l}{0.210$^{+0.019}_{-0.033}$, 0.231$^{+0.017}_{-0.029}$, 0.232 $\pm$ 0.026, 0.232 $\pm$ 0.025} \medskip\\

\hline
\multicolumn{1}{l}{} & \emph{Planet b} & \emph{Planet c} \smallskip\\
Semi-major axis, $a$ [AU]                                 & 0.0762 $\pm$ 0.0022          & 0.1001 $\pm$ 0.0029 \\
Eccentricity, $e$                                         & 0.119$^{+0.082}_{-0.035}$      & 0.095$^{+0.073}_{-0.035}$ \\
Inclination, $i$ [\degree]$^{\bullet}$                     & 88.87$^{\mathrel{+}0.16}_{-0.60}$        & 88.92$^{+0.14}_{-0.41}$$^{\dagger}$ \\
Argument of pericentre, $\omega$ [\degree]                & 179 $\pm$ 52                 & 237$^{+15}_{-68}$ \\
Longitude of the ascending node, $n$ [\degree]$^{\bullet}$ & 180 (fixed)                  & 173.1$^{+2.9}_{-5.6}$  \\
Mean anomaly, $M$ [\degree]                               & 253$^{+61}_{-27}$             & 110$^{+54}_{-34}$ \\
Radius ratio, $R_p/R_\star$$^{\bullet}$                     & 0.07451$^{+0.0014}_{-0.00045}$ & 0.04515 $\pm$ 7.3$\times 10^{-4} $ \\
Planet mass, $M_{p}$ [$M_{\earth}$]                             & 44 $\pm$ 12                & 15.9$^{+7.7}_{-2.8}$  \\
Planet radius, $R_{p}$[$R_{\earth}$]                           & 7.46 $\pm$ 0.76            & 4.51 $\pm$ 0.47 \\
Planet density, $\rho_p$ [$g\;cm^{-3}$]                    & 0.492$^{+0.26}_{-0.092}$       & 0.94$^{+0.46}_{-0.19}$ \\
Planet surface gravity, $\log$\,$g_{p}$ [cgs]              & 2.923$^{+0.058}_{-0.17}$       & 2.952$^{+0.090}_{-0.15}$ \\
$\alpha_1$ [BJD-2,450,000]$^{\ddagger, \bullet}$             & 6813.38356 $\pm$ 4.5$\times 10^{-4}$ & 6817.2732 $\pm$ 0.0013\\
$\alpha_2$ [days]$^{\ddagger, \bullet}$                      & 7.92008 $\pm$ 4.0$\times 10^{-4}$    & 11.9068 $\pm$ 0.0013 \medskip\\

SOPHIE jitter$^{\bullet}$          & 1.15$^{+0.47}_{-0.16}$ & \\
Photometric jitter$^{\bullet,\ast}$ & \multicolumn{2}{l}{1.016 $\pm$ 0.052, 0.717 $\pm$ 0.020, 1.429$^{+0.094}_{-0.041}$, 1.403$^{+0.060}_{-0.12}$} \medskip\\

$q_+ = \frac{M_{p,b}+M_{p,c}}{M_\star}\,\,$$^{\bullet}$ & 0.000198 $\pm$ 4.7 $\times 10^{-5}$  & \\
$q_p =\frac{M_{p,c}}{M_{p,b}}\,\,$$^{\bullet}$         &  0.42 $\pm$ 0.12 & \\
$e_c \cos \omega_c - \frac{a_b}{a_c} e_b \cos \omega_b \,\,$$^{\bullet}$ & 0.0146 $\pm$ 0.0043 &\\
$e_c \cos \omega_c + \frac{a_b}{a_c} e_b \cos \omega_b \,\,$$^{\bullet}$ & -0.168$^{+0.11}_{-0.070}$ &\\
$e_c \sin \omega_c - \frac{a_b}{a_c} e_b \sin \omega_b \,\,$$^{\bullet}$ & -0.067 $\pm$ 0.019 &\\
$e_c \sin \omega_c + \frac{a_b}{a_c} e_b \sin \omega_b \,\,$$^{\bullet}$ & -0.02$^{+0.11}_{-0.18}$ &\\

\hline
\end{tabular}
\begin{list}{}{}
\item $^{\bullet}$ MCMC jump parameter 
\item $^{\dagger}$ reflected with respect to $i = 90^\circ$, the supplementary angle is equally probable.
\item $^{\ast}$ values for K2, NITES, C2PU, and Balesta respectively  made to reduce the number of lines
\item $^{\ddagger}$  $\alpha_1 \equiv t_{\mathrm{ref}} - \frac{\alpha_2}{2\pi}\left(M-E+e\sin{E}\right)$ with $E=2\arctan{\left\{\sqrt{\frac{1-e}{1+e}}\tan{\left[\frac{1}{2}\left(\frac{\pi}{2}-\omega\right)\right]}\right\}}$; $\alpha_2 \equiv \sqrt{\frac{4\pi^2a^{3}}{G M_{\star} }}$. 
\item   $^{\cdot}$ \Msun\ = $1.98842\times 10^{30}\,$kg, \Rsun = $6.95508 \times 10^{8}\,$m,  $M_{\earth} = 5.9736 \times 10^{24}$~kg, $R_{\earth} = 6.378137 \times 10^{6}$~m  
\end{list}
\end{table*}

\subsection{Transit times}
To derive the transit times, we calculate the mid point between the first and fourth transit contact using the \textsc{mercury} dynamic model output. Therefore, our transit time measurements include information on the system architecture and dynamics and as such are better constrained than direct measurement of the transit times in the light curve. However, we assume
 that only the host star and the 2 discussed planets are present in the system. For the case of Kepler-117, it was shown that the precision of the transit times with the photo-dynamic model is three times the precision of a direct fit of the transit times \citep{Almenara2015}.

\begin{figure}
\centering
\includegraphics[width=0.45\textwidth]{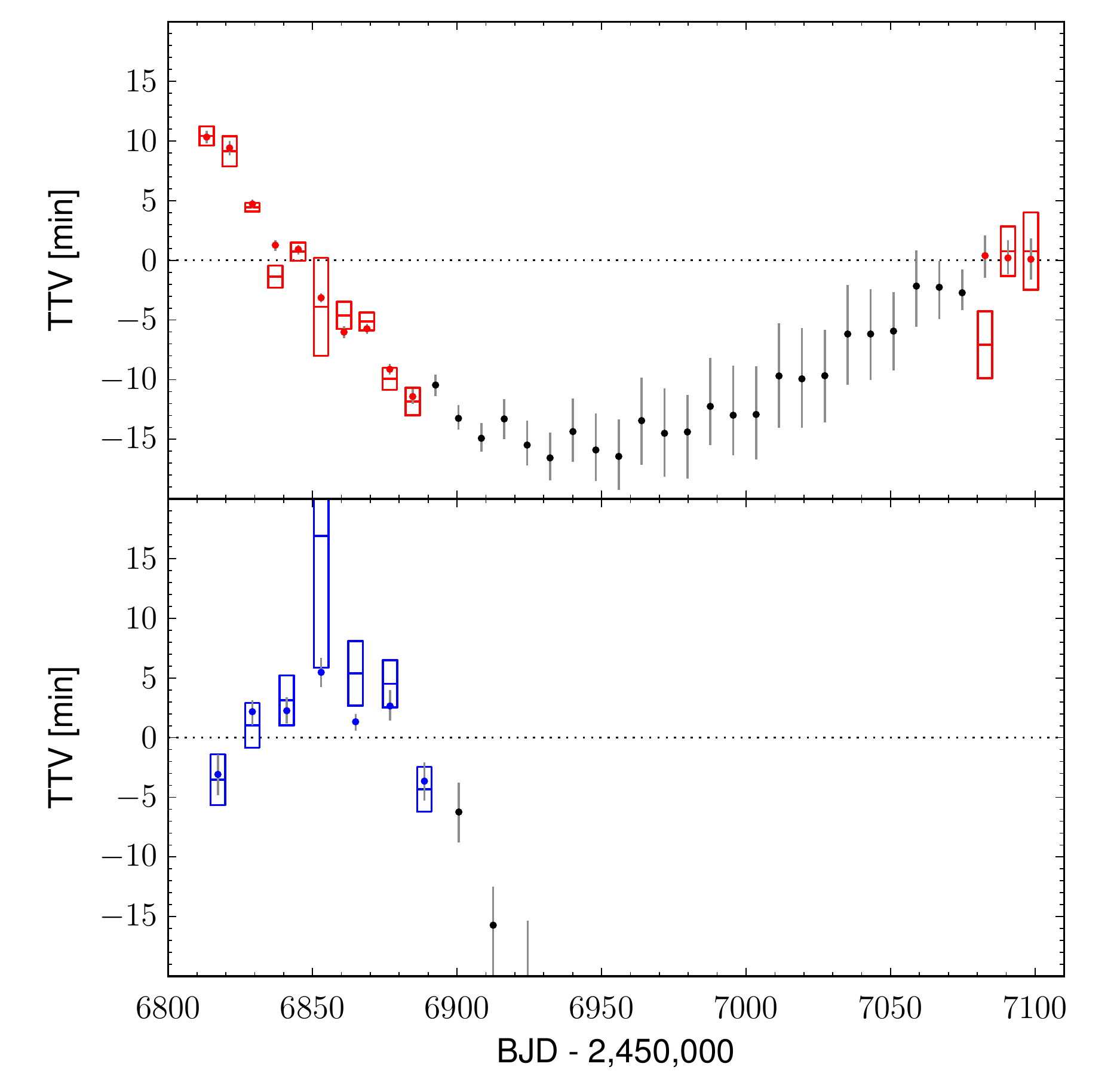}
\caption{Comparison of the TTVs derived by the photo-dynamic model (as circles) with TTVs derived using a standard transit fitting (as boxes with the size of the 1 sigma error) for planet b (top panel) and planet c (bottom panel).  For each planet we use the respective ephemeris derived using the photo-dynamic estimated values of only the observed transits which are marked in red for planet b and in blue for planet c. \label{fig.TTVs} }
\end{figure}

 For comparison, we computed the transit times directly from the K2 light curves using a procedure similar to what is described in \citet{Barros2013}. For each planet the transits were fitted simultaneously ensuring the same transit shape. In Figure~\ref{fig.TTVs}, we compare the derived transit times using the photo-dynamic model and the transit times derived with a standard procedure. To compute the ephemeris we use only the values of the observed transit times derived with the photo-dynamic model. For planet b we derived the ephemeris: $T_b$ (BJD) = $7.921101(69) \times$ Epoch + $2456813.3767(21)$ and for planet c $T_c$ (BJD) = $11.90727(58) \times $Epoch + $2456817.2755(22)$. For each planet the respective and same ephemeris was subtracted from the transit times derived with both methods so that we could directly compare them. 
 
 We find that the difference of the transits times for both methods is less than $3\sigma$ hence the transit times from both methods agree. The higher discrepancy is found for epochs  3 and 34 of planet b (the 4th and 35th datapoint in Figure~\ref{fig.TTVs}).  The transit at epoch 3 shows signs of systematic noise and the transit at epoch 34 is incomplete. It has been shown by \citet{Barros2013} that, in these cases, the errors of the transit times are underestimated, therefore, a difference of  $3\sigma$ is not surprising. Noteworthy,  when analysing the results of the traditional method we find that a linear ephemeris can be rejected at $\sim 100\%\, $ and $\sim 99.9\%\,$ respectively for planet b and planet c. Moreover, for both planets, the periodogram of the transit times shows a peak at the synodic period of the system, albeit not significant.
 
Using our photo-dynamic method, we obtain the double of the precision of the transit times as compared to the traditional method that does not include the dynamical constrains. For the \e350 system the difference increases the significance of the TTVs for planet b and planet c, even in the short duration of the K2 observations, allowing us to better constraint the system architecture. Furthermore, the dynamical constrain reduces the impact of individual transits both with poor normalisation or stellar activity, as long as enough transits are observed. This is because as shown by \citet{Deck2015}, TTVs may contain redundant information on the planetary masses. On the other hand, the TTVs derived with the photo-dynamic model are not pure measurements and depend on the validity of the model assumptions. For example, the existence of other planets in the system that could perturb the observed TTV has to be tested by comparing the photo-dynamic model derived TTVs with those derived directly from the light curve, as we did.

\citet{Armstrong2015} predicted that the resonant timescale of the system is $\sim 5\,$years and hence it is not detectable with the current observation baseline. However, as mentioned above, the TTVs can show a chopping signal at the much shorter synodic timescale. This chopping is clearly visible in Figure~\ref{fig.TTVs}, every 3 orbits of planet b it has a close encounter with planet c that changes its orbit and the transit times. This was also seen in KOI-884 system \citep{Nesvorny2014b,Nesvorny2014}. In the same figure the chopping is also seen for \e350c. Probably the closest encounter of both planets happens near the transit time since planet c and planet b show simultaneous transits during the K2 observations. 

\begin{figure}
\centering
\includegraphics[width=0.45\textwidth]{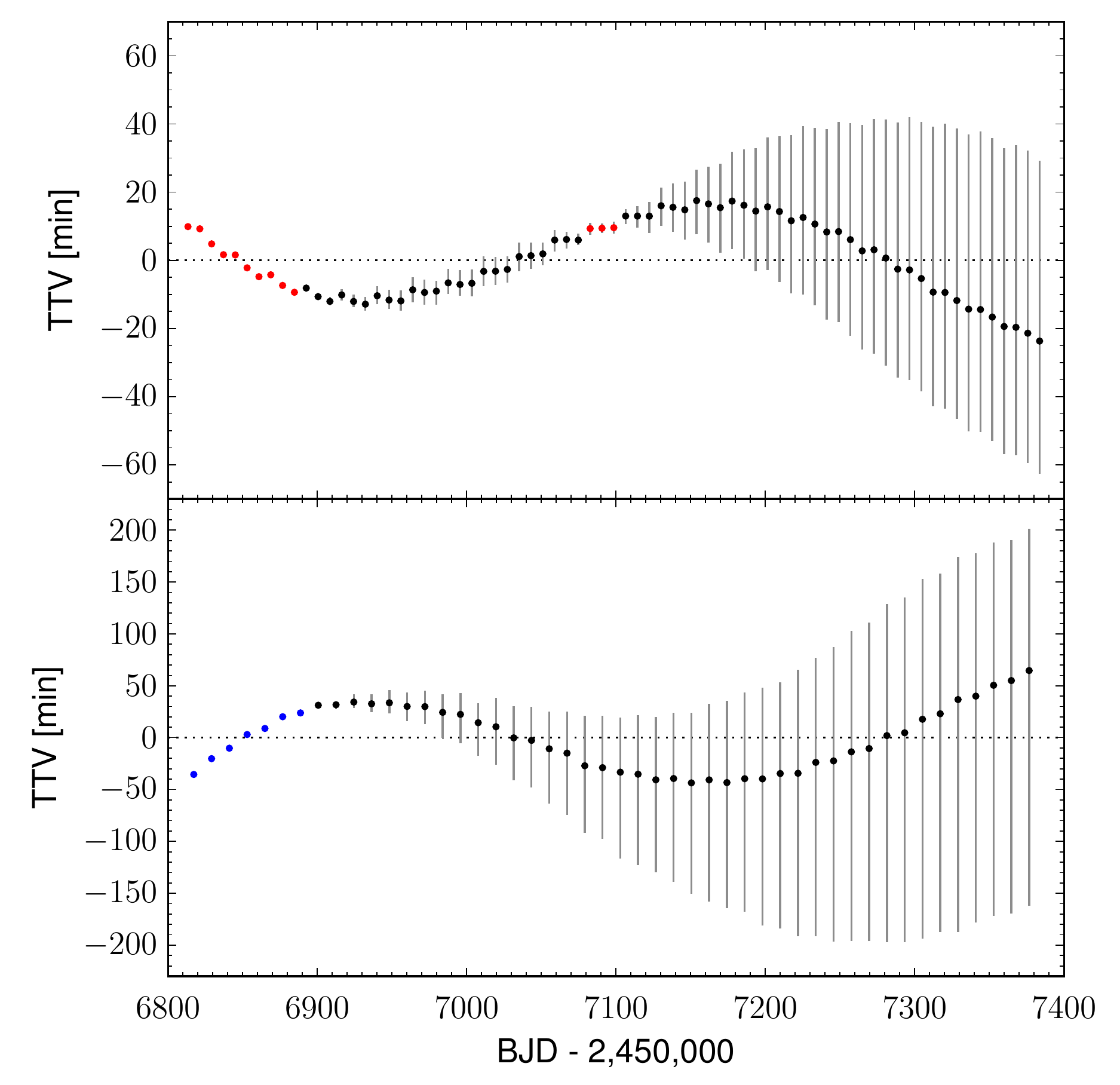}
\caption{Prediction of the TTVs according to the photo-dynamic model until the end of 2015 for planet b (top panel) and planet c (bottom panel). For each planet  we use the respective ephemeris derived from the points plotted. For  \e350b the chopping signal is also visible in this figure. Three transits are nearly on a linear ephemeris and there is an offset from the next three transits due to the conjunctions with the outer planet.   \label{fig.TTVevol}}
\end{figure}

In our case the detection of the chopping signal at the short synodic timescale allows us to directly determine planetary masses. This can be intuitively understood using the equations derived by \citet{Nesvorny2014, Deck2015} although our system might not obey their model assumptions. However, as expected, without the detection of the libration period the orbital eccentricities are poorly constrained. To better constrain the libration period of the system follow-up transits are very important and we encourage further follow-up in the next months to years. To illustrate our uncertainty in the libration period and estimate an ephemeris for the system we evolved 1000 random steps of our MCMC chain till the end of 2015. In Figure~\ref{fig.TTVevol}, we show the model predicted TTVs with the 1 sigma uncertainty derived from the assembly of models. It is clear that we cannot predict the transit times with good accuracy further than 200 days into the future. This is because of the uncertainty in the system parameters and the fact that we do not sample the full libration cycle.  Noteworthy is the different shape of the TTVs of \e350c in Figures~\ref{fig.TTVs} and \ref{fig.TTVevol}. This is due to the different ephemeris (especially the mean period) used to calculate the TTVs. In each case the ephemeris was calculated from different sets of transits times, for Figure~\ref{fig.TTVs} we use only the observed transits while for Figure~\ref{fig.TTVevol} we used all the times presented in that figure. Since the duration of the observations for K2 is too short to sample the resonant timescale, the period measured from K2 observations can be significantly different from the mean period of the system which can only be observationally probed with a much longer time span of the observation.

We found no transit duration variations for the observed transits. Interestingly, the difference on the ascending node of both planets is $6.9^{+2.9}_{-5.6}$, if this is found to be significant it implies orbit precession that leads to a variation of the transit duration. We estimate that by the end of 2017 the transit duration of planet b will increase by $0.23 \pm 0.12\,$ hours.  For planet c we expect a higher change but with higher uncertainty. Further high precision transits are needed to probe the long term evolution of the system and help constrain the system parameters.

\subsection{Model tests}
Transit time measurements can be affected by red noise in the light curve \citep{Barros2013}. These spurious TTVs could lead to an over-fitting of the model and a trapping of the solution. To test this hypothesis we multiplied the errors of the K2 transit light curves by three and repeated the analysis. We find that the errors of the measured transit times increase by a factor of 2-3 but the model is still constrained and we obtain system parameters within $1\sigma$ of the previous results although with uncertainties that are up to 50\% higher.

To further test our model we use the photo-dynamic model described above fitting only the K2 light curve and using neither radial velocity nor ground-based transits nor stellar priors. As expected, the derived parameter distributions are wider, however we still find the best solution in agreement with the previous results. In particular the mass ratios are very well constrained $q_+ $= $0.000159^{+0.000075}_{-0.000018} $ and $q_p$ = $0.481^{+0.24}_{-0.076} $. Because we do not include stellar priors, the scale is not constrained. While stellar density is somewhat constrained by the transits $\rho_{\star} = 2.04^{+1.7}_{-0.49} \rho_\odot$, the stellar radius is unconstrained by the observations and its posterior distribution has the shape of the uniform prior. 

\begin{figure}
\centering
\includegraphics[width=0.45\textwidth]{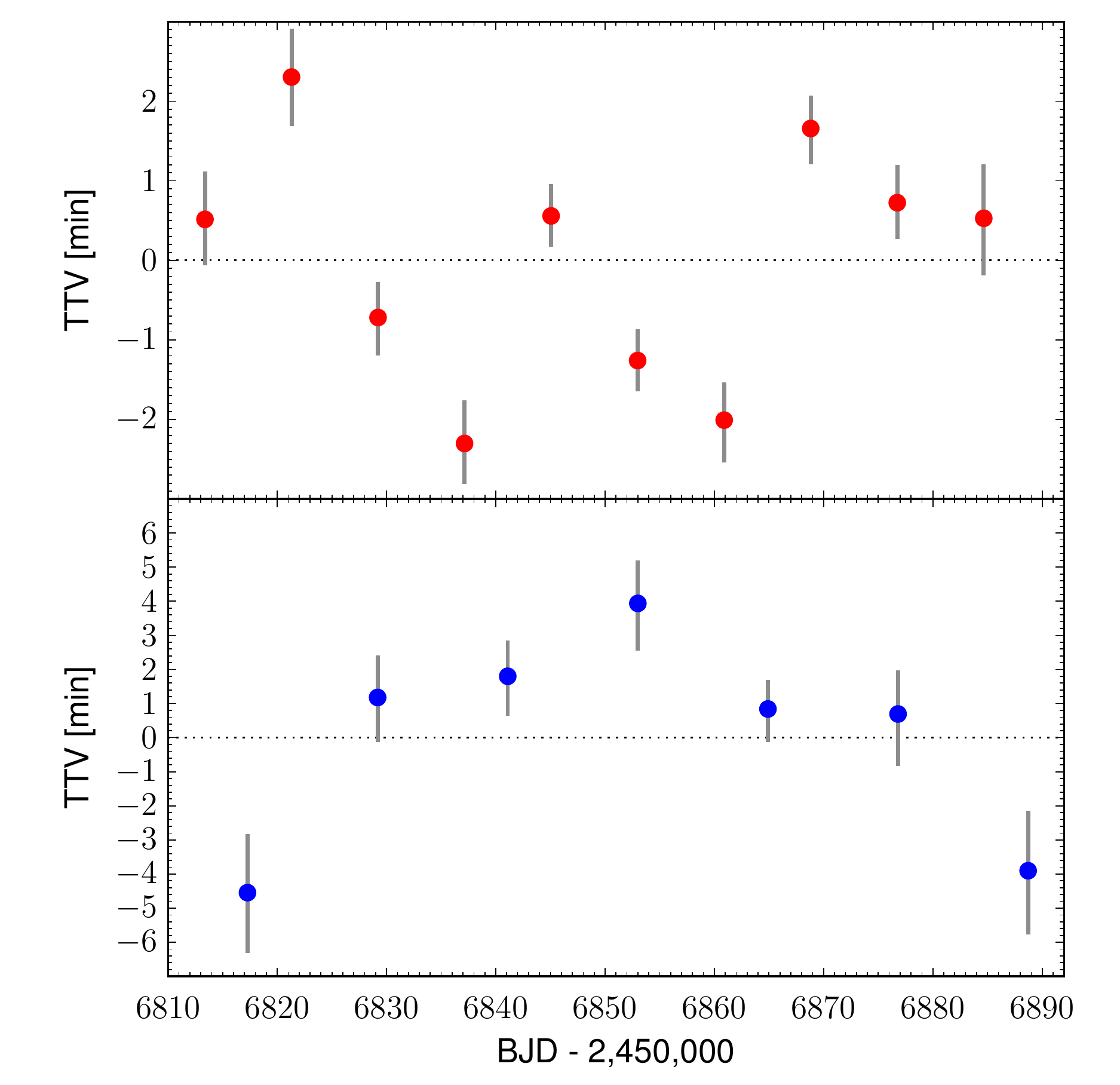}
\caption{Transit times derived from K2 observations with the photo-dynamic model after removing a linear ephemeris. \label{k2ttvs}  }
\end{figure}

 This test using only the K2 light curve predicts the times of the follow-up ground based transits to be $2457082.65858_{-0.094}^{+0.076}$, $2457090.57608_{-0.10}^{+0.082}$, $2457098.49117_{-0.11}^{+0.09}$ respectively for epochs 34, 35 and 36 which are within $1$ sigma of the measured values. Therefore, we conclude that our system solution is robust and it is not significantly affected by spurious TTV due to systematics or spots. The TTVs derived using only the K2 observations are shown in Figure~\ref{k2ttvs} where the chopping is clearly visible. So this method will be useful for short duration observations like K2, TESS and CHEOPS.

\section{Discussion}

\begin{figure}
\centering
\includegraphics[width=0.45\textwidth, angle=90]{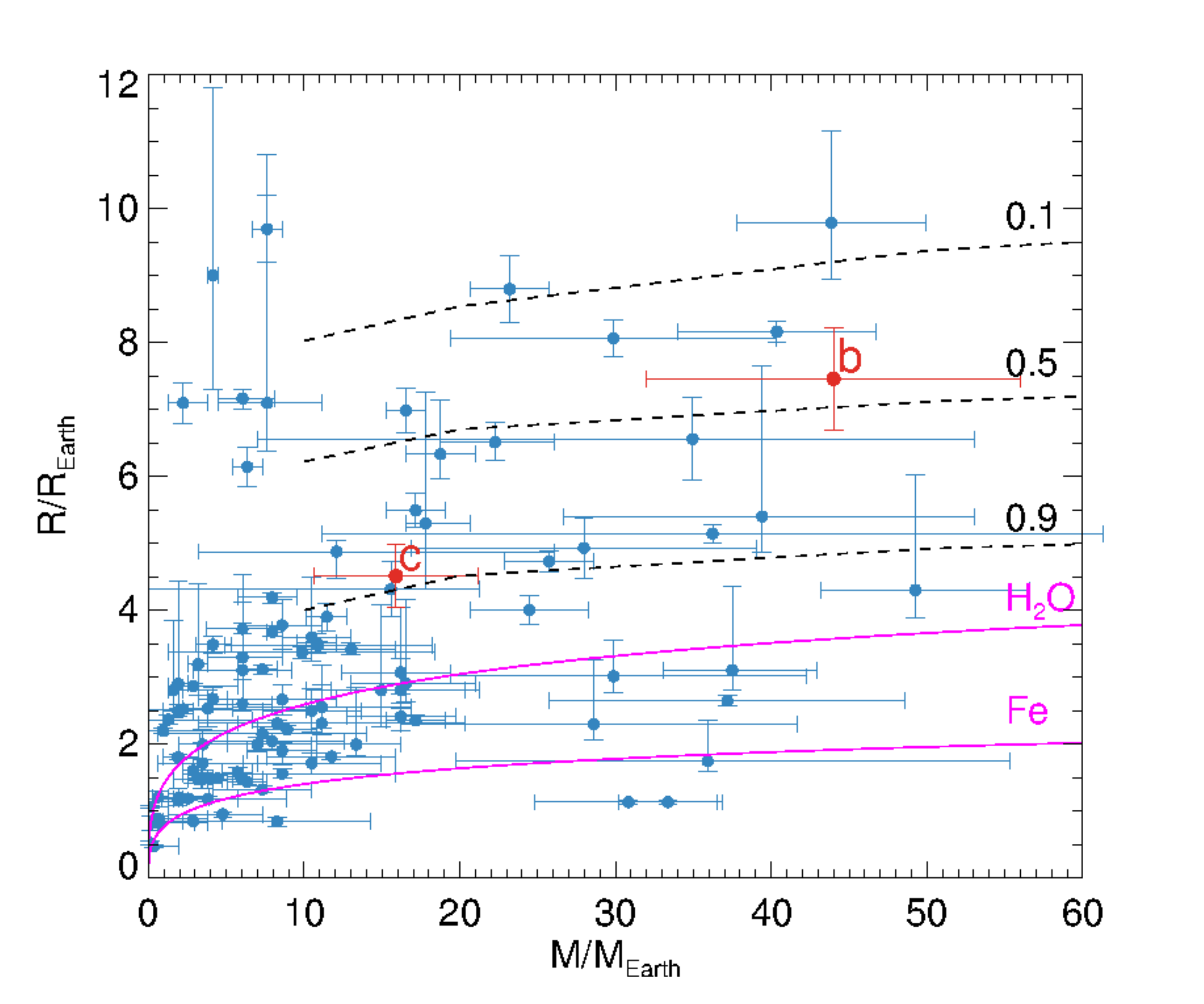}
\caption{ Mass-radius relationship for different bulk composition of planets. In magenta solid lines we show the models of  \citet{Zeng2013} for solid planets with a composition of pure iron and pure water. In black dashed lines we show the models of \citet{Baraffe2008} for gaseous planets with heavy material enrichments of Z = 0.1,0.5 and 0.9 assuming and an age of 5~Gyr.\label{mr} We superimposed the known planets in this mass-radius range and the position of the planet \e350b and \e350c marked with a respective red letter.  }
\end{figure}

We present a photo-dynamic analysis of the \e350 system discovered in the C1 campaign of K2. This system is composed of a slightly metal-rich K dwarf with $ M_*$= $0.949 \pm 0.077 \,$ \Msun\ and $  R_* $= $0.913 \pm 0.094  $  \Rsun and two transiting planets very close to 3:2 MMR. \e350b has a mass,  $M_b $= $44 \pm 12\,$ \mearth, a radius $R_b$ = $7.46 \pm 0.76\, $\rearth\ and an orbital period  $P_b \sim 7.92\,$days and  \e350c has a mass,  $M_c$ = $15.9^{+7.7}_{-2.8}\,$ \mearth,  a radius $R_c$ = $4.51 \pm 0.47\, $\rearth\ and an orbital period,  $P_c \sim 11.91\,$days. \e350c is similar to Uranus. The radius derived for both planets is in agreement with those derived previously by \citet{Armstrong2015}. However, our analysis allow a much better constraint on the mass the planets.
In Figure~\ref{mr} we show the position of \e350b and \e350c in the mass-radius diagram compared with known planets with M < 50~\mearth and R < 10~\rearth . 
 In the same figure we plot the theoretical models for solid planets with a composition of pure iron and pure water \citep{Zeng2013}. We also plot the models of \citet{Baraffe2008} which apply to planets with gaseous envelopes  and different heavy material enrichments assuming and an age of 5~Gyr. Clearly,  \e350b and \e350c are less dense than a  pure water/ice sphere, therefore they have a significant gas envelope. \e350c probably has a high fraction of heavy elements and hence, a core while \e350b probably has less heavy elements which results in a lower density.

We present follow-up transit photometry and spectroscopic observations that together with the K2 observations allow us to uniquely characterise the system.
We also show that even using only the K2 photometry, we can derived the mass ratios of the system and find the same unique solution. 
This is due to the detection of the chopping short timescale TTVs in both planets produced by the orbital conjunctions between the planets. In our case the non-detection of the full libration curve leads to a poor constraint on the orbital eccentricity but does not prevent finding a unique solution. We predict a libration period of $\sim 1.5$ years but further observations are needed to confirm it. We thus encourage further follow-up observations to sample the full TTV curve and derive the eccentricity. This will give further insight into the evolution and formation of this system \citep{kley&nelson2012}, which is specially interesting since it is one of the closest systems to the 3:2 MMR known.

The predicted radial velocity amplitude for planet b is $14.4 \pm 4.2\,$\ms and for planet c is $4.9^{+2.0}_{-1.0}\,$\ms, below the precision of SOPHIE for this target. However, they can be measured with any of the HARPS instruments since the target is visible from both the Canary Islands and La Silla. Comparing the TTV predicted mass with the one measured by radial velocities will be important to validate the TTV mass determination method and probe the existence of additional companions. \citet{Weiss2014} state that amongst their sample of 65 exoplanets smaller than 4 Earth radii, the masses determined by TTVs are statistically lower than masses estimated by RVs. An example of this discrepancy is
KOI-94-d  \citep{Borucki2011} whose mass is $106 \pm 11 $\mearth\  \citep{Weiss2013}  according to the RV analysis, whereas the TTV analysis gives $52.1 \pm 7$ \mearth\  
 \citep{Masuda2013}.  However, for KOI-142b (Kepler-88b) the TTV derived masses are in agreement with the RVs derived masses \citep{Nesvorny2013, Barros2014}. Understanding the discrepancies is very important to allow direct comparison between the TTV and RV estimated masses. Furthermore, if the reflex motion of the star is measured it is possible to derive masses and radius for the system bodies without using stellar models \citep{Almenara2015}.

In traditional transit timing variations analysis of multi-planetary systems, the individual TTVs are first derived from transit fitting and later modelled using n-body dynamic simulations to constrain planetary masses. We show that fitting simultaneously the transit light curves with the system dynamics (photo-dynamical model) increases the precision of the TTV measurements and helps constrain the system architecture. Without this increase in precision we would not detect short period TTVs (chopping) and the mass ratios would be very poorly constrained \citep{Armstrong2015}. Our method is a powerful tool for the analysis of the short duration K2 data and future observations  with CHEOPS and TESS.  Applying a photo-dynamic model will help characterise a higher number of systems and will lead to better understanding of the evolution of near-resonant systems.

\section*{Acknowledgements}
We thank the staff at Haute-Provence Observatory. SCCB acknowledges support by grants 98761 by CNES and the Funda\c c\~ao para a Ci\^encia e a Tecnologia  (FCT) through the Investigador FCT Contract No. IF/01312/2014. This work was also supported by FCT through the research grant UID/FIS/04434/2013. JMA acknowledges funding from the European Research Council under the ERC Grant Agreement n. 337591-ExTrA. OD acknowledges support by the CNES grant 124378. A.S. is supported by the European Union under a Marie Curie Intra-European Fellowship for Career Development with reference FP7-PEOPLE-2013-IEF, number 627202. We thank the referee for his suggestions that improved the manuscript.

\bibliographystyle{mnras} 
\bibliography{susana}









\bsp	
\label{lastpage}
\end{document}